# Trajectory-dependent electronic excitations by light and heavy ions around and below the Bohr velocity

S. Lohmann [a], R. Holeňák [a], D. Primetzhofer [a]

[a] *Department of Physics and Astronomy, Uppsala University, Box 516, 751 20 Uppsala, Sweden*

**Abstract**

We present experiments demonstrating trajectory-dependent electronic excitations at low ion velocities, where ions are expected to primarily interact with delocalized valence electrons. The energy loss of $H^+$, $H_2^+$, $He^+$, $B^+$, $N^+$, $Ne^+$, $^{28/29}Si^+$ and $Ar^+$ in self-supporting silicon membranes was analysed along channelled and random trajectories in a transmission approach. For all ions, we observe a difference in electronic stopping dependent on crystal orientation. For heavier ions, the energy-loss difference between channelling and random geometry is generally found more pronounced, and, in contrast to protons, increases for decreasing ion energy. Due to the inefficiency of core-electron excitations at employed ion velocities, we explain these results by reionization events occurring in close collisions of ions with target atoms, which are heavily suppressed for channelled trajectories. These processes result in trajectory-dependent mean charge states, which strongly affects the energy loss. The strength of the effect seems to exhibit a $Z_1$ oscillation with an observed minimum for Ne. We, furthermore, demonstrate that the simplicity of our experimental geometry leads to results that can serve as excellent benchmark systems for dynamic calculations of the electronic systems of solids using time-dependent density functional theory.

## 1. Introduction

Energy transferred by energetic charged particles to matter governs a number of astrophysical phenomena [1] and leads to radiation damage in extreme environments [2]. Detailed understanding of the energy-deposition mechanisms allows not only for predicting the mentioned effects but also for using this process for a number of scientific and technological applications ranging from hadron therapy for cancer treatment [3] to materials characterisation and modification [4]. Particularly for semiconductors, the use of ion beam irradiations is widely employed to manipulate material properties via implantation or controlled defect creation [5,6].

The energy deposition of ions is commonly denoted by the stopping power $S$, which is defined as the average energy loss per unit path length, i.e. $S = -dE/dx$. Stopping has been the subject of extensive research for decades both by modelling and experiments [7]. Many experimental studies determine $S$, in accordance with its definition, as an effective average along the ion trajectory, often by employing amorphous or polycrystalline samples. Likewise, several theoretical approaches have successfully predicted $S$ without even taking the atomic or electronic structure of the target material into account [8,9].

The individual energy-transfer events from ions to target constituents are, however, well-known to be impact parameter dependent [10]. So, despite the successes of averaging approaches, the specific nature of energy deposition at the nanoscale can only be understood by going beyond them. From an experimentalist's point of view, an impact-parameter selection can be achieved by employing samples with a long-range order, i.e. single-crystals. When an ion travels through a crystal with its direction of motion closely aligned with a crystal axis, it becomes subject to the channelling effect. Channelling denotes the small-angle scattering of ions by collective potentials of atomic strings at large impact parameters, leading to an oscillating motion confined to the channel [11]. At high ion energies, channelling is well studied, and results in enhanced ion ranges, i.e. significantly reduced stopping, along the channel due to the suppression of core-electron excitations in close collisions (see e.g. the reviews by Gemmell [12] and Cohen and Dauvergne [13]).

At lower ion velocities, low meaning around and below the Bohr velocity $v_0$, however, ion-solid interactions behave differently than at high ones, also affecting the impact-parameter dependence. Core-electron excitations become much less efficient, and the interactions with target valence electrons become non-adiabatic. A direct consequence is dynamic screening of the charge of the penetrating ion which affects its scattering probabilities. Moreover, charge-exchange events and the formation of molecular orbitals [14] directly, and indirectly via altering the mean charge state of the ion, affect the energy loss, and again the scattering potential [15]. Additionally, the structure of the target density of states has to be taken into account [16], although the influence on electronic interactions is often complex [17,18]. In particular, ions heavier than protons might significantly perturb the electronic system of the target altering the response of the system [19,20].

With the development of time-dependent density functional theory (TD-DFT), a modelling tool to adequately describe these dynamic processes has become available [21]. For the sake of computational simplicity, these simulations are often done for well-defined lattices, i.e. mimic the transmission of ions through single-crystals. Several ion-target combinations have been studied this way, e.g. H in Si, Ge and graphite [22–24], Si in Si [25,26], Ni in Ni [27] and W in W [28]; however, a comparison to adequate experimental data, i.e. data obtained under exact same conditions using well-defined samples, is often missing.

In a previous study we have measured the energy loss of slow protons and He ions through self-supporting Si(100) nanomembranes both for channelling and random conditions [29]. Results for protons agree well with previous studies at higher energies, and exhibit an increase of the difference between channelled and random trajectories with ion energy due to increasing contributions of core-electron excitations for close collisions. For He ions, we observed a reverse trend, i.e. an increase of the difference between random and channelled energy loss for lower ion velocities. We explained this behaviour by collision-induced charge-exchange events resulting in a higher mean charge state of the ion, at energies where the equilibrium charge state approaches zero. These charge exchange processes lead to a direct energy loss due to electron promotion but more importantly increases the electronic stopping along random trajectories. The more complex electronic structure of heavier ions implying the possibility for transitions involving multiple electrons plus a larger range of available charge states, can be expected to lead to even more pronounced differences. In addition, the electronic stopping of slow channelled ions has been predicted to oscillate with the projectile atomic number $Z_1$ by many theoretical and experimental studies [30–32]. This oscillation has also been observed specifically in single-crystalline Si [33–36], even though only one of these works is experimental, and employed samples are not well-defined. For amorphous targets, $Z_1$ oscillations have been observed e.g. in C [37], but the amplitude of the oscillation seems to be much reduced compared to channelling geometries. Similarly, a recent study on polycrystalline TiN showed, if at all, a very weak oscillation with $Z_1$, in large discrepancy to theoretical predictions by DFT [38]. Experiments that study $Z_1$ oscillations for the same system and under identical conditions for random and channelling geometry have, however, not performed in the named studies. Such a direct comparison between these two cases is expected to clarify the trajectory-dependence of electronic excitations and, specifically, $Z_1$ oscillations.

We have, for these reasons, conducted a follow-up study to [29] investigating the energy loss of $B^+$, $N^+$, $Ne^+$, $^{28/29}Si^+$ and $Ar^+$ ions with velocities $< v_0$ in well-defined and characterised Si(100) nanomembranes. We measure 2D distributions of particles transmitted through the self-supporting foil targets and compare the energy deposition in channelling and random geometries to access the impact-parameter dependence of electronic excitation processes. We furthermore compare our results on self-irradiated silicon to available TD-DFT calculations.

## 2. Experimental methods

All experiments were performed using single-crystalline, self-supporting Si(100) foils (Norcada Inc. "UberFlat" silicone membranes) with nominal thicknesses between 50 and 200 nm as samples. Rutherford backscattering spectrometry (RBS) and time-of-flight elastic recoil detection analysis (ToF-ERDA) were employed for characterising sample areal density and purity. For the RBS measurements, 2 MeV $He^+$ ions provided by the 5 MV 15SDH-2 Pelletron Tandem accelerator at Uppsala University were used as probes and backscattered ions were detected at 170° scattering angle with a passivated implanted planar silicon detector. To avoid channelling, incidence angles between 5° and 30° were chosen, and small rotations around this set value were performed by the acquisition software to further randomize the alignment. For all measurements the beam current, as measured on a gold reference, was kept below 5 nA to avoid damaging the foils. The thinnest samples (thickness 53 nm) were measured relative to a bulk gold sample. Hereby, the beam was directed onto the reference for 15 s, then onto the sample of interest for 30 s, then again onto the reference and so on. The number of incident particles per solid angle was consequently determined for the reference and used to evaluate the Si(100) spectrum. Analysis to obtain areal densities was performed with help of the SIMNRA software [39]. The uncertainty of obtained results is expected to be better than 6 %. For samples measured by the relative method described, this uncertainty increases to about 7.5 %.

ToF-ERDA measurements were likewise performed with the Tandem accelerator using 36 MeV I$^{8+}$ ions and detecting recoiling target atoms at 45° recoil angle [40]. Energy and flight time of recoils is measured in coincidence allowing for a separation of light contaminants from the Si matrix. Results indicate high purity in the bulk and mainly, H, C and O contamination in accordance with expectations of a thin surface oxide and spurious contamination with hydrocarbons.

All energy loss measurements were conducted with the time-of-flight medium energy ion scattering (ToF-MEIS) system at Uppsala University [41,42]. Ion beams are provided by a Danfysik implanter platform and available energies range from 20 to 350 keV for singly charged ions. Pulsing of the beam is performed by an electrostatic chopper combined with a gating pulse resulting in pulse widths down to 1 to 3 ns. Several sets of horizontal and vertical slits allow for a restriction of the beam cross section to well below (1 x 1) mm$^2$ and a beam divergence significantly better than 0.056° (for comparison, the smallest used membranes have an area of (3 x 3) mm$^2$). The current incident on the sample can consequently be reduced to 2 – 3 fA only. The base pressure in the experimental chamber is found to be below 1 x 10$^{-8}$ mbar. No specific sample cleaning has been performed on-site for the presented experiments.

We measured the energy loss of H$^+$, H$_2^+$, He$^+$, B$^+$, N$^+$, Ne$^+$, $^{28/29}$Si$^+$ and Ar$^+$ ions through self-supporting Si(100). Three different sample thicknesses were used: 53 nm, 135 nm and 200 nm, as obtained from the areal densities measured with RBS. Ions are transmitted through the foil, and detected 290 mm behind the sample with a position-sensitive microchannel plate (MCP) detector (DLD120 from RoentDek [43]). The position is determined with the help of two perpendicular delay lines, and the energy of transmitted particles is measured via their flight time. The circular detector has a diameter of 120 mm, corresponding to deflection angles ±11.5°, and covers a solid angle of 0.13 sr. Samples are mounted to a 6-axis goniometer allowing not only for precise positioning but also for studying different beam-crystal alignments.

The ToF transmission approach together with the large size of our detector allows for 3D mapping of obtained results, i.e. a plot with information on energy and transmitted intensity available for every pixel. Details of the evaluation procedure and different available contrast modes are presented in a separate publication [44]. Figure 1 shows the energy of transmitted ions as a function of position on the detector. In both cases, $^{29}$Si$^+$ ions with initial energies of 100 keV are employed as probes and Si(100) foils with 53 nm thickness as samples. The projected position of the incident beam is indicated by the small white circle (the innermost in a). The position-dependent mean energy after transmission is calculated for bins sized (0.5 x 0.5) mm$^2$. In a, the sample is positioned in such a way that the [100] crystal axis is aligned parallel to the incident beam. In this geometry, the large majority of ions is channelled, and arrives at the detector at small scattering angles around the incident beam position. Some ions experience scattering by larger angles, and are detected in outward regions of the detector. These ions are further influenced by the crystalline structure on their outward trajectory, i.e. they are subject to blocking and planar channelling. These effects also allow for real-space images of the crystal structure. The colour contrast gives the position-dependent energies of ions after transmission, which can simply be transformed into energy loss values by subtraction from the initial beam energy. Ions exhibit the highest detected energy, i.e. the lowest energy loss, along the channelled trajectory. Energy loss increases for larger scattering angles, however, planar channelling leads to energies lying between the axial channelling energy in the detector centre and energies measured between the planes. This effect results in the star-like pattern visible in the figure. Note, that the expected energy loss from scattering by an angle equivalent to the half-opening angle of the detector including increased trajectory length is expected to increase the energy loss by about 5%.

For the measurement plotted in Fig. 1b, the sample was rotated by $\theta_x = 6°$ around the *x* axis and by $\theta_y = 12°$ around the *y* axis. Then, the incident beam is no longer aligned with any low-index crystal axis, and we call this geometry "random". Otherwise, experimental conditions are the same as for the channelled incidence. The intensity of transmitted ions does not peak distinctively around the projected position of the incident beam indicating that more ions undergo scattering at larger angles. The energy distribution likewise exhibits a smaller range with a significantly lower maximum than in the channelled case. Lower energy loss in the crystal planes and high-index channels near the incident-beam direction can still be observed. Details of the energy loss will be further discussed in the following sections.

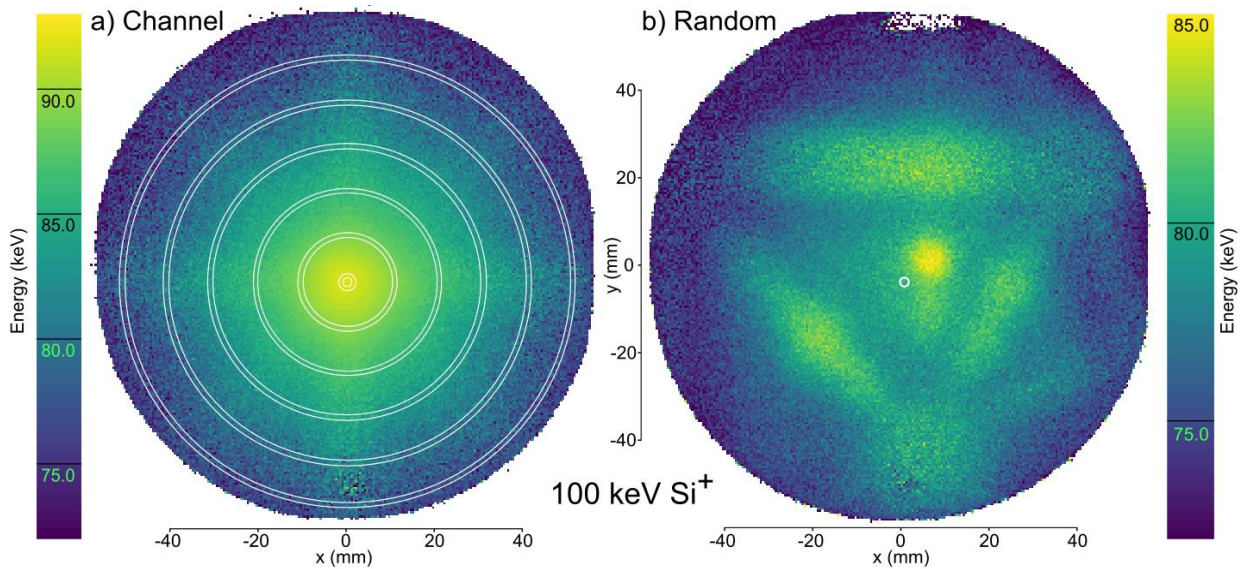

Figure 1: Position-dependent energy of $^{29}Si^+$ ions after transmission through a 53 nm thick, self-supporting Si(100) foil. The initial ion energy is 100 keV, and the initial beam position is indicated by the small white circle (the innermost one in a). a) the incident beam is aligned with the principal [100] crystal axis (channelling geometry). b) the sample has been rotated by $\theta_x = 6°$ around the *x* axis and by $\theta_y = 12°$ around the *y* axis arriving in a (pseudo-)random position, in which the beam is not aligned with any low-index channel or plane. The annuli drawn into a) visualise regions of interest evaluated in Fig. 2.

The plots presented in Fig. 1 indicate that energy loss differences between trajectories can be assessed in two different ways: (i) different regions of interest (ROIs) on the detector corresponding to different scattering angles can be evaluated and (ii) the sample can be rotated to compare different incidence conditions. Both methods are employed in this work. Small circular ROIs around the initial beam position (as e.g. indicated by the innermost circle drawn into Fig. 1) can be selected to study rather straight trajectories. In order to study exclusively the electronic energy loss contribution, the nuclear energy loss has to be considered. Even assuming suppressed nuclear stopping due to a strong trajectory selectivity in the present approach [45,46], the contribution is not fully negligible for the heaviest projectiles and a random geometry. We, therefore, perform simulations of transmission experiments using the Monte-Carlo package TRBS [47] to estimate the contribution from elastic collisions. TRBS allows for modification of the Firsov screening length in the employed Thomas–Fermi-Molière potential [48]. By setting this correction factor to 1 % of the standard value, trajectories virtually without elastic collisions are simulated and can be compared to simulations resembling the experiment, i.e. including potential nuclear losses. The angular range of simulated transmitted particles was chosen to be 0 – 0.8° to consider straight trajectories only, but still achieve sufficiently high statistics. After subtraction of the expected nuclear energy loss, the electronic stopping power can be determined from the measured energy loss employing numerical integration and assuming a $E^{1/2}$ dependence of the stopping power. For the narrow energy intervals, the inaccuracies introduced by a potentially different scaling are expected to be smaller than other experimental uncertainties.

## 3. Results

We first evaluate energy loss as a function of detection angle for 100 keV $^{29}$Si$^+$ ions transmitted through 53 nm Si(100), i.e. the measurements already presented in Fig. 1. To this aim, only ions arriving in narrow annuli centred around the incident beam position are selected. These ROIs are also drawn into Fig. 1a. Resulting energy loss spectra are depicted in Fig. 2. Results for channelling and random incident geometry are compared and shown in Fig. 2a and b, respectively. All curves are normalised to their respective maximum value for easier comparison.

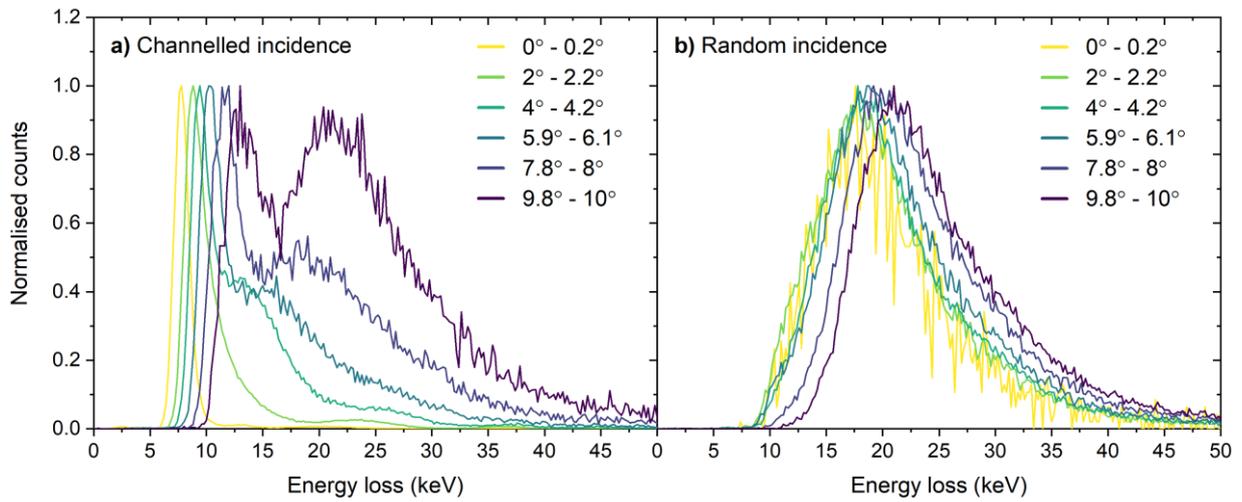

Figure 2: Energy loss spectra for different deflection angles recorded by transmitting $^{29}$Si$^+$ ions with initial energy 100 keV through a self-supporting 53 nm thick Si(100) sample. The spectra in a) (channelled incidence) and b) (random incidence) correspond to the spatial distributions depicted in Fig. 1a and b, respectively. Note that each figure shows data from one single measurement. The different curves are obtained by evaluating different regions of interest on the detector (annuli drawn into Fig. 1). The angular ranges given in the legend correspond to the widths of these annuli. All curves are normalised to their maximum value.

Ions incident in channelled geometry and exiting at very small deflection angles (the yellow curve in Fig. 2a) exhibit a very narrow energy distribution and the lowest energy loss of all shown cases. For larger angles, the energy loss gradually increases, and the distribution broadens. From around 4° (the dark green curve) onwards, the energy loss distribution shows two distinct peaks, which separate more clearly for larger deflection angles. For random incidence, the energy distribution is very broad even for the innermost ROI. The shape of the distribution does not change with deflection angle and only slightly shifts towards higher energy loss for larger angles. A comparison between the darkest curves in Fig. 2a) and b) shows that the high-energy loss peak for the channelled-incidence spectrum is located at the same position as the random-incidence peak for an equivalent deflection angle of 10°.

A significant difference in the energy loss observed along random and channelled trajectories for the shown example becomes already apparent from Figures 1 and 2. To study this observation more in detail and compare with our previous results for protons and He ions, we evaluate the energy loss along rather straight trajectories ending in circular ROIs with 1 mm radius around the initial beam position (corresponding to deflection angles $\pm 0.2°$) for channelled ($\Delta E_{ch}$) and random incidence ($\Delta E_r$). Results for all studied ions, in the form of the ratio $\Delta E_{ch}/\Delta E_r$ as a function of initial ion velocity (given in atomic units a.u.), are compiled in Fig. 3. Error bars include the time resolution of the respective measurement and the uncertainty in the flight path caused by the finite size of the beam spot and the evaluated ROI. Note that the ratio is shown for the measured energy loss, not stopping power. Due to the large difference between $\Delta E_{ch}$ and $\Delta E_r$ and the non-linear energy dependence of the stopping power slightly different values can be expected for the latter.

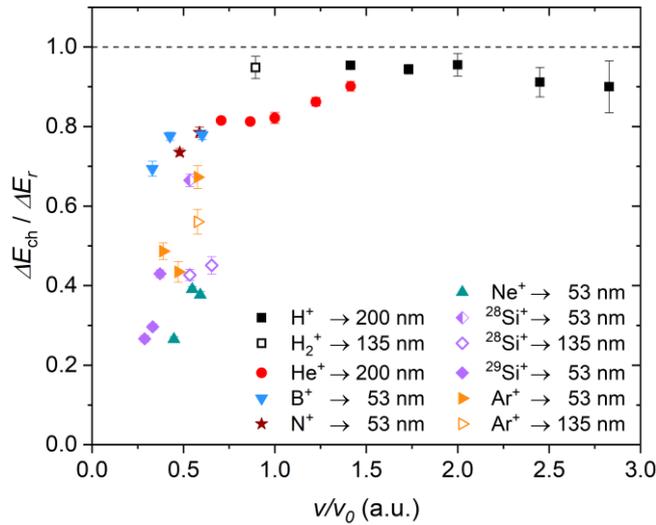

Figure 3: Comparison between energy loss along channelled $\Delta E_{ch}$ and random trajectories $\Delta E_r$ denoted by the fraction $\Delta E_{ch}/\Delta E_r$ for H$^+$, H$_2^+$, He$^+$, N$^+$, Ne$^+$, $^{28/29}$Si$^+$ and Ar$^+$ ions transmitted through self-supporting Si(100) membranes. Data for light ions from [29] (H$^+$ and He$^+$) is shown next to ratios obtained for heavier ions with velocities well below $v_0$. Energy loss is evaluated along rather straight trajectories by selecting only ions that are detected in small regions of interest around the incident beam position (for details see text).

All ions heavier than protons exhibit a similar behaviour – a strong deviation of $\Delta E_{ch}/\Delta E_r$ from unity and mostly a decrease of the ratio for lower velocities. Whereas for the lowest studied He velocity the channelled energy loss reaches about 0.82 of the random one, much larger differences are observed for heavier ions. As an example, $\Delta E_{ch}$ at the lowest studied velocity corresponding to 60 keV Si reaches only 0.27 of $\Delta E_r$. The difference does not gradually increase with $Z_1$, however. Instead an apparent $Z_1$ oscillation is observed with the $\Delta E_{ch}/\Delta E_r$ ratio first decreasing from He over N to Ne and then rising again over Si to Ar, when comparing data recorded at similar ion velocities. A complete quantification of this effect is hampered by large deviations between different data points both for Si and Ar though.

## 4. Discussion and conclusions

The differences in energy loss for random incidence but different deflection angles can be well explained by different path length through the sample and additional kinematic losses in a single-scattering event with the respective scattering angle. For example, SIMNRA simulations for Si in Si exhibit a difference of 3 keV between a scattering angle of close to 0° and 10°, in good agreement with our observations. The double peak structure for channelled incidence needs an additional explanation though. The comparison with the random incidence curve at the largest studied deflection angle indicates that the peak structure featuring a high energy loss consists of ions that have travelled on completely random trajectories, i.e. have been dechannelled very close to the sample entry point. Ions with lower energy loss must have been dechannelled later, i.e. their trajectories comprise channelled parts of various lengths. The clear separation between the two peaks indicates that the dechannelling probability is not constant over the channel length. We interpret this feature as an increased probability for dechannelling close to both surfaces, in particular in both surface oxides. Note that even a surface reconstruction, as commonly observed for clean Si(100), would lead to a similar process [49], although the magnitude of the effect is expected to differ between the two cases. In other words, a large fraction of ions detected at larger deflection angles is dechannelled when entering the crystal (high energy loss) and another large fraction when exiting (low energy loss). Figure 1a) also indicates the occurrence of planar channelling, which would constitute another trajectory with an expected energy loss between the channelled and the random one. The broadening of the distribution suggests a mix of different trajectories ending in the same ROI on the detector. This interpretation, together with the high contrasts of observed blocking patterns and the results obtained from ToF-ERDA measurements, also confirms the high purity of and lack of lattice distortions in the crystalline bulk of employed samples.

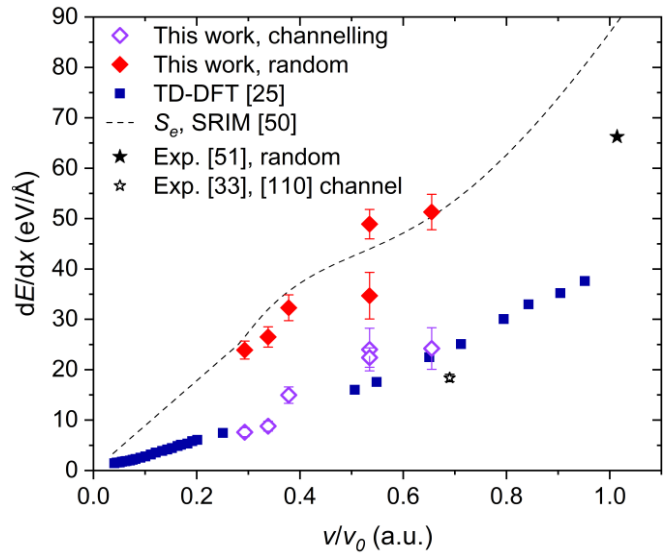

Figure 4: Comparison of experimental stopping powers of Si in Si with data from TD-DFT. The energy loss of $^{28/29}Si^+$ was measured in transmission through self-supporting Si(100) nanomembranes. For details on employed isotope-target thickness combinations see Fig. 3. Purple open diamonds represent data measured in channelling geometry whereas for red filled diamonds the sample was rotated into a random alignment with the beam. The dashed line gives the prediction from SRIM [50] and the black asterisk, measured by Arstila [51], denotes the data point at the lowest energy available in the IAEA stopping power data base at the time of writing. The black open asterisk denotes experimental data from [33], measured in Si(110), however. TD-DFT calculations (dark blue squares) performed by Lim et al. [25] also model the passage of Si ions along the Si [100] axis.

Finally, we want to compare our results to predictions from ab-initio calculations. Specifically, we compare our data set for Si projectiles with TD-DFT calculations from [25]. The results are shown in Fig. 4. Purple open diamonds denote electronic stopping powers obtained in channelling geometry, whereas red filled diamonds correspond to a random beam-crystal alignment. The electronic stopping power was determined according to the procedure described in Section 2 including the subtraction of nuclear energy loss for random geometry. For the case at hand, elastic losses were found to contribute between 23 % (lowest energy) and 4 % (highest energy). The figure, furthermore, shows predictions from SRIM (dashed line) and an experimental data point from [51] (obtained in pseudo-random geometry) corresponding to the lowest energy data available in the IAEA stopping power data base at the time of writing. The open asterisk denotes electronic stopping measured along the [110] axis from [33]. Our data measured in channelling geometry and the TD-DFT calculation that also models the passage of Si ions along the [100] axis of a Si crystal show excellent agreement. Electronic stopping powers along random trajectories are well predicted by SRIM. The scatter of datapoints at a velocity of 0.55 in Fig. 4 is expected to result from an inaccuracy or time evolution in the beam alignment during experiments but still features good agreement with predictions on average. Note, that even though the relative discrepancy between channelling and random, as visible in Fig. 3, decreases, the absolute one increases.

For slow heavy ions, critical angles for axial and planar channelling are expected to be much larger than for H and He. Thus, finding a completely random alignment of the beam for the employed samples is experimentally not straightforward. As an example, in Fig. 1b) channelling in a high-index channel close to the incident beam position is observed (the high-energy region to the top-right of the white circle). In this context, as mentioned above, in Fig. 4, intercomparing the random data points with each other and the SRIM data suggests that the low electronic stopping at 0.55 a.u. might not correspond to a true random stopping, but has been measured over a partially channelled trajectory. The ratios presented in Fig. 3 for ions heavier than N can, thus, be perceived as upper limits rather than absolute values.

Since core-electron excitations at employed ion velocities are inefficient (except for protons at the highest employed velocities) and the comparison with Monte Carlo calculations shows that even for heavy projectiles, nuclear losses only comprise a minor energy-loss channel, other mechanisms have to be the cause for the observed difference between channelled and random trajectories. In [29] we propose that in the case of He ions, repeated capture and loss processes of electrons lead to the measured higher energy loss along random trajectories. Whereas the neutralisation of ions via Auger processes happens at any impact parameter, relevant reionisation mechanisms can only occur in close collisions that are strongly suppressed for channelled trajectories. For one-electron processes, promotion of electron levels occurring at small impact parameters is known to reduce the He kinetic energy by about 20 eV [52]. Two-electron processes require more energy, but are expected to be much rarer [53]. In both cases, this direct effect is not sufficient to explain the much larger observed differences even for many charge-exchange cycles. Therefore, we suggest that the major contribution to the increased energy loss along random trajectories is an increased mean charge state of the ion with an additional minor contribution from energy dissipated directly in the excitation process.

The reported results for heavier ions can be attributed to similar mechanisms as described for He. The much lower $\Delta E_{\text{ch}}/\Delta E_r$ values are expected to be caused by dynamic processes involving several electrons and higher possible charge states. For self-irradiated Si, TD-DFT calculations by Lee et al. show indeed a strong dependence of the electronic stopping of channelled Si ions on the initial ion charge state [26]. Furthermore, different equilibrium charge states for channelling and off-channelling projectiles are reported, thus, corroborating our explanation.

The observed apparent oscillatory behaviour indicates a different dependence on $Z_1$ of the electronic stopping power along channelled and random trajectories. Whereas our available energy range does not permit measurements for the exact same velocity for all projectiles, we qualitatively compare the oscillation of the electronic stopping power with $Z_1$ for the data points closest to 0.6 a.u. We observe an oscillation both for random and channelling geometry, however, the latter is significantly more pronounced. The minimum for $Z_1$ = 10 (Ne) agrees well with previous studies on ions channelled in Si that report a pronounced minimum in the stopping around 10 or 11. Analysing our results in the context of existing literature [33,37,38], it can be concluded that $Z_1$-oscillations of electronic stopping are a consequence of the binary collisions between the non-excited projectile with the valence and conduction electrons of the target, in accordance with predictions from theory [54]. Projectile excitations, which become important along random trajectories, seem to severely counteract this effect. With increasing equilibrium charge state, both the amplitude of the $Z_1$-oscillation but also the relative discrepancy between the energy loss in channelling and random trajectory is diminishing. The latter discrepancy, however, is expected to increase again at elevated energies when the contribution of core-electron excitations to the energy loss becomes substantial.

## 5. Summary and outlook

We measured the energy loss of B, N, Ne, Si and Ar ions in Si(100) along channelled and random trajectories in transmission. We found a significantly reduced energy loss in channelling geometry even for these low employed velocities below the Bohr velocity. For deflected ions, this trajectory dependence leads to a noticeable splitting of the energy distribution of an initially monoenergetic channelled beam. We explained this behaviour by an increased dechannelling probability at both surfaces.

In contrast to higher velocities, core-electron excitations are expected to be suppressed for all geometries. We, therefore, proposed the observed difference in energy loss to be routed in repeated electron-capture and -loss processes induced in close encounters with target nuclei available only for random trajectories. This mechanism is expected to excite the projectile and significantly raise the ion mean charge state, which leads to higher electronic stopping power. The strength of this effect depends on the projectile electronic structure, and we observed a drastically less pronounced oscillation with $Z_1$ for random geometry compared to channelled trajectories. Provided that sample surfaces are sufficiently clean, the mean charge state distribution can be accessed by measuring ion exit charge states. Experiments of this kind for different crystal orientations could, therefore, help to study the observed phenomena further.

We also showed that the simplicity of our experimental approach together with the high quality of employed samples can provide excellent benchmark data for stopping powers calculated with TD-DFT for ion velocities below the Bohr velocity where experimental data is scarce. Nevertheless, TD-DFT codes will need to include charge-exchange mechanisms to accurately predict stopping powers even for polycrystalline and amorphous materials.

**Acknowledgements**

Support of accelerator operation by the Swedish Research Council VR-RFI (Contract No. 2017-00646_9) and the Swedish Foundation for Strategic Research (Contract No. RIF14-0053) is gratefully acknowledged.